# Evolution of thermomagnetic instability along a superconducting wire.

G.L. Dorofeev, E.P. Krasnoperov, Yu.D. Kuroedov, V.S. Vyatkin
RRC "Kurchatov Institute", Moscow, Russia

*Abstract*

Propagation of a magnetic perturbation along a monofilament superconducting wire under thermomagnetic instability was investigated. In NbTi wire at temperature T=4.2K and magnetic field B=0.5-0.7T the propagation velocity is found v = 2.5 – 3.5 km/s. In NbZr the corresponding value v= 9 km/s.

It is known that a hard superconductor will be in unstable state when magnetized in a magnetic field. With random increasing temperature the screening currents decrease and magnetic flux penetrates into the sample (flux jump). A review of theoretical and experimental works concerning the conditions for critical state stability can be found in reference [1]. The thermomagnetic instability occurs when the magnetic field difference inside and outside the superconductor exceeds a critical value $B_k$. For NbTi alloys at liquid helium temperature (T=4.2K), the magnetic flux jumps into cylindrical sample becomes possible for magnetic field higher than $B_l \geq 0.3$ T [2]. The time evolution of a local thermomagnetic instability and the relaxation time for the instability have been investigated extensively ([1,3] and references therein). The magnetic flux jumps are usually observed at the changing external magnetic field. Magnetic perturbation produces addition heating. It results early magnetic flux jumps occur. As far as we are aware, the propagation of magnetic perturbation along a wire has not been studied.

The purpose of the present work is to report kinetics of perturbation propagation (the wave of magnetic flux penetration) along a superconducting wire in absence of transport current. The NbTi and NbZr monofilament wires of diameter 0.08 mm to 0.24 mm were used. Experiments were carried out at liquid helium temperature. The sample was disposed inside of a long superconducting solenoid. The top part of Fig.1 shows the schematic drawing of a sample geometry. Superconducting solenoid is not shown. Detection coils C1, C2...C6 were fixed on a wire. The distance between adjacent coils was 3 to 5 mm. The initiator coil (denoted as "IC" at the left side of the wire) produces thermal or magnetic pulse that destroys superconductivity under this coil. In normal phase the magnetic field is uniform along radius. Because everywhere in a sample (except into "IC") the magnetic field is not uniform (according Bean's model the magnetic field decreases linearly in a radial direction of a wire), the magnetic field penetration wave has possibility to propagate from "IC" along a wire. When the wave front crosses a detection coil, electric voltage pulse (U) appears on the coil and oscilloscope records this pulse. This one enables to investigate the propagation of magnetic perturbation.

Typical oscillograms, U vs. time, are presented in a Fig. 1. The upper curve corresponds to an initial normal phase excitation after zero field cooling process followed by magnetization of the wire. The first pulse corresponds to the electrical disturbance produced by the "IC". The next six pulses are produced by the magnetic flux that consiquently penetrates in detection coils. The polarity of the pulses depends on an external magnetic field direction. The known distances between coils and the measured time differences between the pulses allows one to determine velocity of flux penetration wave -V. The accuracy of the velocity measurement is about ±5%. The velocity of perturbation propagation is constant along the sample. Since the wire length is much longer than the excitation area (IC size) we can to say that this process is steady state

(V=const). It is important to point out that in the case of fixed temperature and field the next normal phase excitations do not produce signals in detection coils, as shown in the lower diagram in Fig.1. The absence of the signals can be explained that a uniform field distribution is produced. After magnetic penetration waves passed the fields inside and outside the wire are the same. In this case flux jump can not appear. To create a metastable state again the wire had to be heated over a critical temperature, followed by zero field cooling and magnetized.

We have to note that during perturbation propagation the electric resistance of the wire outside the "IC" is much less than the normal state resistance.

Fig. 2 shows the field dependence of the propagation velocity for NbTi wires of diameter 0.08, 0.12, and 0.16 mm. The field dependence shows a threshold character, where the excitation does not propagate for the external field less than $B_l = 0.3T$. The value $B_l$ corresponds to the maximum temperature jump on the sample surface during a magnetic flux jump [2]. Our method determines the lower border of flux jump stability as a local excitation propagation to the large distances, much times more than the excitation area.

The propagation velocity has a maximum in the field region of 0.5 to 0.8 T. For wire of diameter 0.16 mm the largest velocity is V=3.5 km/s. If the field is increased further above the velocity maximum, the propagation velocity reduces and abruptly vanishes (V=0) at field $B_f$, corresponding to the upper limit for thermomagnetic instability. It is easy to estimate the value $B_f$ that increases with increasing wire diameter. For NbTi wire of diameter 0.12 mm the thermomagnetic wave does not propagate for $B_f >0.85T$. It is known that the thermomagnetic instability disappears in a large magnetic field. This is due to the fact that, with a reduction of the critical current density at a high field, the field difference between the inside and outside of wire is also reduced (see [1] and references there). It is clear that thermomagnetic instabilities do not propagate, if the critical current density is lower than $J_c < B_l /R$, where R is radius of wire. With a simple approximation for magnetic field dependence of the critical current density $J_c =\alpha/B$, where $\alpha$ is a constant, the upper limit of a stable state $B_f =\alpha R/ B_l$ increases with a wire radius, in agreement with experiment.

In addition to NbTi wires, the NbZr wires with and without brass covering (diameter 0.33 and 0.24 mm, respectively) were investigated. The field dependence of the wave velocity has the same character like NbTi wire, but for NbZr wire without covering the velocity maximum exceeds 9 km/s. In case of NbZr wire with brass covering the wave velocity is about three times less.

It is obvious a high velocity of thermomagnetic instability evolution and a threshold character of a velocity (fig. 2) have the same nature as a magnetic flux jumps [1,2], but to this time quantitative side of this problem is not solved.

In summary the steady state propagation of magnetic perturbation wave along a monofilament superconducting wires magnetized in a parallel magnetic field has been observed. In NbTi wires at temperature T = 4.2 K and external magnetic field B=0.5 to 0.8T the propagation velocity reaches to V = 2.0 to 3.5 km/s. The highest velocity V = 9 km/s was attained in an NbZr wire.

The authors thank Academician V. E. Fortov for bringing the present problem to their attention and Dr. M.Senba for assistance with proofreading and comments.

**Figure Captions:**

Figure 1. The oscillogram of voltage U on measuring coils vs. time for NbTi wire of diameter 0.16 mm. On the top part of figure the measuring coils (C1 – C6) around a wire and an initiation coil ("IC") are drawn.

Figure 2. Magnetic field dependence of the propagation velocity of magnetic flux penetration wave along superconducting NbTi wires of diameters: 0.08 mm (■); 0.12 mm (Δ) and 0.16 mm (*).

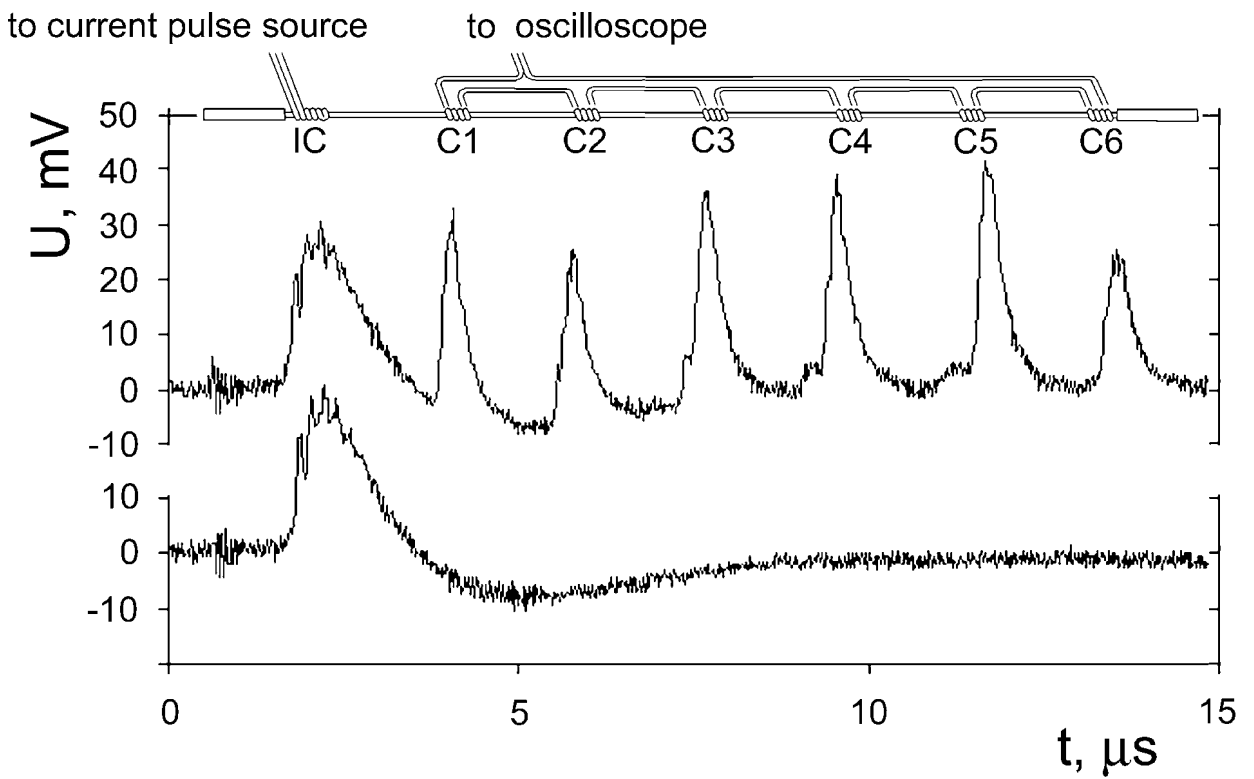

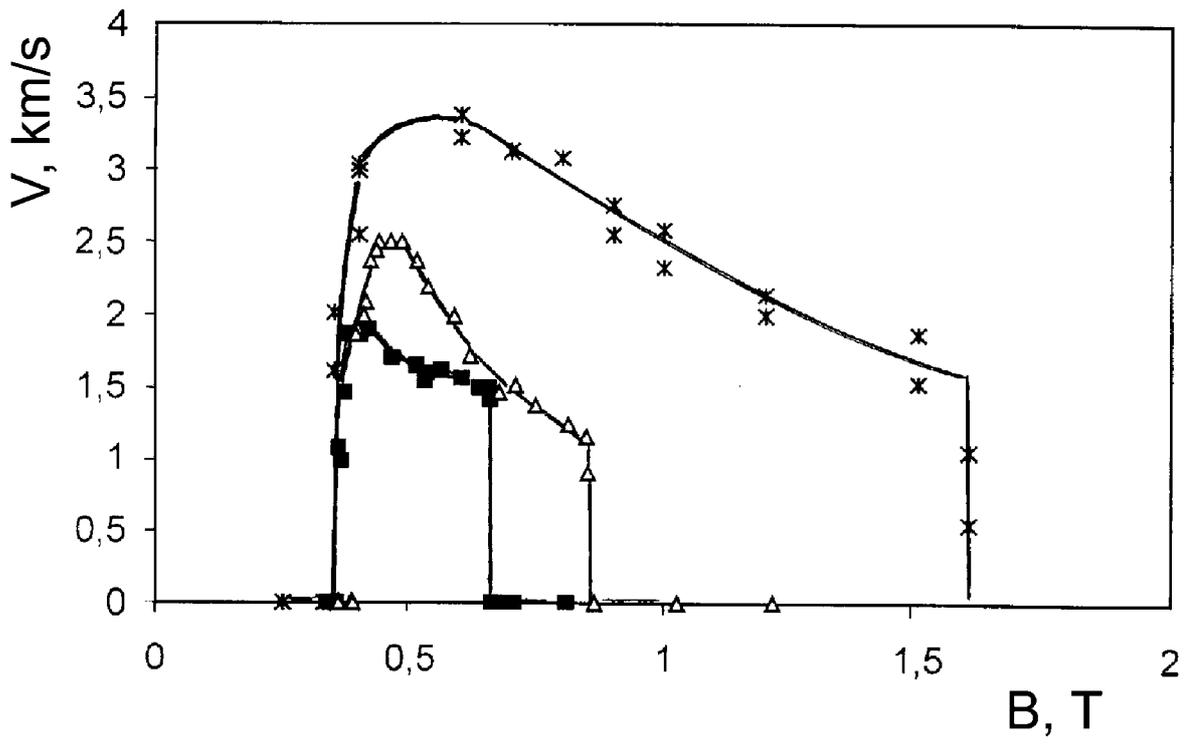